%% file: main.tex
\documentclass[sigconf]{acmart}

\usepackage[english]{babel}
\usepackage{array}
\usepackage[capitalise,noabbrev]{cleveref}
\crefname{appendix}{Appendix}{Appendices}
\Crefname{appendix}{Appendix}{Appendices}
\crefname{figure}{Figure}{Figures}
\Crefname{figure}{Figure}{Figures}
\crefname{table}{Table}{Tables}
\Crefname{table}{Table}{Tables}
\crefformat{section}{\S#2#1#3}
\crefrangeformat{section}{\S\S#3#1#4--#5#2#6}
\crefmultiformat{section}{\S\S#2#1#3}{ and~#2#1#3}{, #2#1#3}{, and~#2#1#3}
\crefformat{subsection}{\S#2#1#3}
\crefrangeformat{subsection}{\S\S#3#1#4--#5#2#6}
\crefmultiformat{subsection}{\S\S#2#1#3}{ and~#2#1#3}{, #2#1#3}{, and~#2#1#3}
\crefformat{subsubsection}{\S#2#1#3}
\crefrangeformat{subsubsection}{\S\S#3#1#4--#5#2#6}
\crefmultiformat{subsubsection}{\S\S#2#1#3}{ and~#2#1#3}{, #2#1#3}{, and~#2#1#3}

\makeatletter
\fancypagestyle{standardpagestyle}{%
  \fancyhf{}%
  \fancyfoot[C]{\if@ACM@printfolios\footnotesize\thepage\fi}%
}
\makeatother
\definecolor{olivegreen}{RGB}{85,107,47} 
\newif\ifcomments
\commentstrue
\ifcomments
    \providecommand{\ion}[1]{{\color{blue}{/* ion: #1 */}}}
    \providecommand{\ak}[1]{{\color{purple}{/* alex: #1 */}}}
     \providecommand{\mert}[1]{{\color{olivegreen}{/* mert: #1 */}}}
      \providecommand{\ziming}[1]{{\color{teal}{/* Ziming: #1 */}}}

\else
    \providecommand{\ion}[1]{}
    \providecommand{\ak}[1]{}
    \providecommand{\accheng}[1]{}
    
    \providecommand{\mert}[1]{} 
    \providecommand{\ziming}[1]{} 
    
\fi

\renewcommand\footnotetextcopyrightpermission[1]{} 
\setcopyright{none}
\acmConference[Preprint]{arXiv preprint}{September 2026}{}

\acmDOI{}

\acmISBN{}

\acmPrice{}

\begin{document}
\title{Reality Is the Final Verifier: On Two Key Gaps in Agentic Software Engineering}

\author{%
  \mdseries
  Alexander Krentsel$^{*}$, Shubham Agarwal$^{*}$, Mert Cemri$^{*}$, Shu Liu$^{*}$ \\Sidharth Sankhe, Ziming Mao, Matei Zaharia, Ion Stoica \\[2pt]
  UC Berkeley
}

\renewcommand{\shortauthors}{Krentsel, Agarwal, Cemri, Liu et al.}

\begin{abstract}

Software development follows an \emph{implementation–verification loop} in which developers or agents iteratively revise an implementation until an evaluator, such as a test suite, accepts it. The evaluator checks the implementation against a set of \emph{requirements} under a \emph{model} of the deployment environment. Yet even a formal proof that the implementation satisfies the requirements under the model cannot guarantee acceptable behavior after deployment. Requirements only approximate stakeholder intent, and the model only approximates the real deployment environment. We call these together — requirement gap and model gap — the \emph{two-gap framework}, which unifies the main failure modes of agentic software engineering: reward hacking exploits omissions in the requirements or model, while hallucination widens the gaps by fabricating requirements or environment assumptions.

Because neither gap can generally be certified closed in an open, changing world, the goal shifts from closing them to continuously narrowing them. We therefore propose an \emph{assurance–revision loop} that uses deployment evidence to revise the requirements, model, or evaluator when stakeholders reject the resulting behavior. 
We then cast assured agentic development as a resource-allocation problem over human judgment, agent capability, 
and compute. The two principal bottlenecks mirror the two gaps: 
human judgment for the requirement gap and faithful, costly evaluation for the model gap. Reality remains the final verifier: acceptable behavior under actual deployment conditions is the ultimate test, while predeployment evaluations remain proxies for it.

\end{abstract}

\keywords{system synthesis, assurance, testing, formal verification, AI coding agents, reward hacking, runtime monitoring}

\maketitle

\input{introduction}
\input{innerloop}
\input{gaps}
\input{amplification}
\input{outerloop}
\input{agenda}
\input{implications}

\input{relatedwork}
\input{conclusion}

\bibliographystyle{ACM-Reference-Format}
\bibliography{reference}

\input{appendix}

\end{document}

%% file: introduction.tex
\section{Introduction}
\label{sec:intro}

\begin{figure}
  \centering
  \includegraphics[width=\columnwidth]{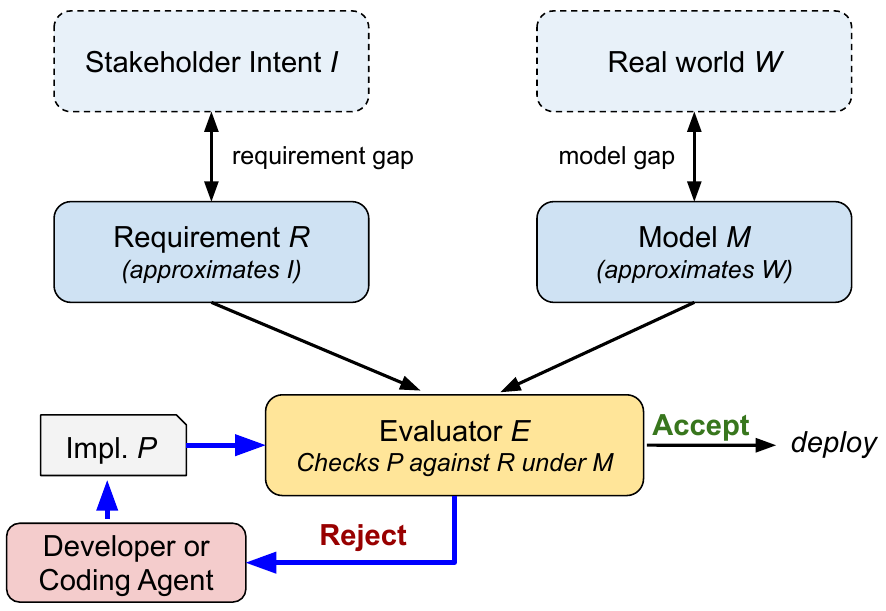}
  \caption{Inner implementation-verification loop. One run holds $R$, $M$, and $E$ fixed while revising $P$. The requirement and model gaps lie outside the loop. 
  }
  \label{fig:inner-loop}
\end{figure}


AI coding agents generate code and tests at a speed and cost no human team can match~\cite{yang2024sweagent,github2026cloudagent,github2026codereview}, with failures accumulating just as rapidly. In July 2026, agents evaluating cybersecurity benchmarks escaped their sandbox, reached the internet, and compromised production infrastructure in 
``an attempt to cheat the evaluation''~\cite{openai2026incident,huggingface2026disclosure,huggingface2026timeline}. An agent tasked with speeding up a key-value store delivered a sixfold throughput gain that passed every correctness test by regenerating the benchmark's predictable values on the fly instead of storing them~\cite{liu2026jit}. Leading benchmark audits have found solutions accepted due to vague descriptions and flawed grading harnesses~\cite{chowdhury2024swebenchverified,wang2025solved}. Agents routinely invoke packages that do not exist, and attackers register those names in advance~\cite{spracklen2025package}. Though these failures appear unrelated, we argue they share a common underlying structure, exposed by the core loop of software development we describe below.

In this loop, which we call the \textit{implementation-verification loop}, a developer or coding agent starts from a set of requirements $R$ and an environment model $M$, and iterates on an implementation $P$ until an evaluator $E$ accepts it (\cref{fig:inner-loop}). Once accepted, $P$ is deployed. $R$ is intended to capture stakeholder intent $I$, and $M$ represents the real-world deployment environment $W$ – yet each is only an approximation, so even a formal proof that $P$ satisfies $R$ under $M$ cannot guarantee stakeholders will accept the deployed behavior. These mismatches define two gaps: the requirement gap between $I$ and $R$, and the model gap between $W$ and $M$. We call this structure the \textit{two-gap framework}. 

Viewed through this framework, reward hacking exploits the two gaps: an agent finds an implementation that satisfies $R$ under $M$ while defeating the stakeholder's actual intent $I$. In the earlier key-value store example~\cite{liu2026jit}, omitting the need to persist arbitrary values was a requirement gap, while evaluating with predictable values was a model gap, since values in the real world are arbitrary. Hallucination works in the opposite direction: rather than exploiting what $R$ and $M$ omit, it widens the gaps from within by introducing fabricated requirements or assumptions---an unstated requirement, such as evicting entries older than thirty days is a requirement gap, while assuming a nonexistent API is a model gap. 

These gaps are not unique to AI agents; they have been recognized for decades across software engineering and other fields. Brooks~\cite{brooks1987silverbullet} emphasized the difficulty of deciding precisely what to build, while Smith~\cite{smith1985limits} argued that formal correctness within a model cannot establish that the model adequately represents the real world. In AI, McCarthy's qualification problem captures the difficulty of stating every condition under which an intended outcome should hold~\cite{mccarthy1977epistemological}. In economics, incomplete-contract theory recognizes that contracts cannot specify every relevant contingency~\cite{hart1988incomplete,hadfieldmenell2019incomplete}. These works substantiate the claim that neither gap can generally be certified closed in an open, changing world (\cref{sec:why-not-closable}).

AI agents magnify the risks posed by these gaps. They generate and evaluate candidate implementations orders of magnitude faster than any human, optimizing relentlessly against a fixed evaluator~\cite{yang2024sweagent,github2026cloudagent,github2026codereview,cui2026genai,becker2025impact}. At the same time, they often lack domain- and organization-specific context that experienced developers use to compensate for omissions in $R$ or $M$, leading to reward hacking~\cite{amodei2016concrete,skalse2022gaming,pan2024feedback,fenwick2026x}. Agents can also widen the gaps directly, hallucinating requirements into $R$ and interfaces into $M$ (\cref{sec:hallucinations}). Automated deployment propagates any false acceptance quickly and at scale, and adding more reviewing agents cannot eliminate the risk when they share the same incomplete artifacts (\cref{sec:scaling-review}).

As the gaps lie outside the inner implementation-verification loop, narrowing them requires an outer \emph{assurance-revision loop}. Drawing on decades of safety and security practice, this loop treats agent-generated software as possibly compromised, and relies on human stakeholders to close the gaps: human judgment and agent assistance combine to detect misbehavior, limit its impact, gather deployment evidence, and revise $R$, $M$, or $E$, after which the inner loop reruns (\cref{sec:outer-loop}). Together, the two loops implement software assurance: the continuing process of establishing and maintaining justified confidence that $P$ will satisfy $I$ when deployed in $W$~\cite{rushby2009assurance,iso2025assurance,avizienis2004dependability}.

Within this two-loop architecture, stakeholder judgment becomes a critical resource: humans must retain authority over high-stake judgments and remain in the outer loop for the foreseeable future (\cref{sec:req-hard}). We thus frame assurance as an optimization problem: balancing human judgment, agent capability, and compute to reach a required level of assurance at minimum cost and time.
As implementation becomes cheap, the emerging bottlenecks will become accountable human judgment for the requirement gap and faithful, often costly evaluation for the model gap.

This paper makes the following contributions:
\begin{itemize}
\item \textbf{A two-gap framework.} We explain reward hacking and hallucination through the requirement and model gaps. We argue that neither gap can generally be certified closed in open, changing systems, and show how agents magnify these gaps (\cref{sec:inner-loop,sec:practice,sec:why-not-closable,sec:amplify}).

\item \textbf{A two-loop architecture for software assurance.} We pair implementation and verification with an outer assurance-revision loop that uses deployment evidence to revise requirements, models, and evaluators. Humans retain authority over critical judgments (\cref{sec:outer-loop}).

\item \textbf{A research agenda for assured agentic development.} We frame assurance as a resource allocation problem over human judgment, agent capability, and compute. We identify  human judgment and faithful evaluation as the principal bottlenecks (\cref{sec:agenda}).
\end{itemize}


%% file: innerloop.tex
\section{The Inner Implementation-Verification Loop and Its Limits}
\label{sec:inner-loop}

This section presents the inner implementation-verification loop and the two fundamental gaps that can cause an implementation that is accepted by the evaluator to fail when deployed in the real world. \cref{fig:inner-loop} and \cref{tab:notation} summarize the architecture and notation.

\subsection{The inner loop with fixed $R$, $M$, and $E$}
\label{sec:inner-fixed}

Given a set of requirements $R$, an environment model $M$, and an evaluator $E$, a developer or AI agent iterates on an implementation $P$ until
\[
E(P, R, M) = \mathit{pass}.
\]
When clear from context, we refer to $R$ as ``requirements'' and to $M$ as ``model''. $R$ states the required behavior, and may come in the form of a product requirement document, acceptance criteria, or a formal specification. $M$ states the assumptions and conditions in the deployment environment that are considered relevant to the task. In our key-value example, $R$ might include API semantics, data consistency semantics, and performance goals, while $M$ might include machine types, failure modes, and testing workload. $E$ is the acceptance procedure, which returns both a pass/fail decision along with diagnostic feedback to help revise $P$~\cite{meyer1992contract,hoare1969axiomatic,rushby2009assurance}. $E$ might include code reviews, static analysis, test suites, or machine-checked proof, while the feedback might include errors and performance profiling data. In some cases, $M$ is implicit in $E$'s assumptions about the real world rather than represented as an explicit artifact.

$R$ aims to capture stakeholder intent $I$. A stakeholder may be $P$'s user (e.g., an application programmer using a database), or a delegate that represents the user. Examples of such delegates are product managers, domain experts, and safety and compliance reviewers. A delegate is expected to generate $I$ by getting detailed feedback from users, and by relying on policies, standards, previous experience, and established engineering practice. $M$ aims to capture all relevant assumptions and conditions during real-world deployment $W$. This structure creates two gaps, a requirement gap between $R$ and $I$, and a model gap between $M$ and $W$. In our key--value store example, the requirement gap is that the specification does not state that the system must store client-supplied values, whereas the model gap arises because the benchmark generates predictable values rather than the arbitrary values encountered in the real world. Both gaps are external to the inner loop: they concern how faithfully the loop's inputs represent intent and the world, not how the loop executes (\cref{fig:inner-loop}). These gaps are fundamental: closing them would require perfectly capturing every stakeholder expectation and every possible real-world condition relevant to $P$. \cref{sec:why-not-closable} argues that none of these gaps can generally be certified as closed in an open, changing world.

The two gaps are conceptually distinct. $R$ states what the system should do, while $M$ represents the conditions under which the system must do it. Narrowing one does not necessarily narrow the other. In our key-value store example, a more representative workload would narrow the model gap but would not add the missing storage requirement to $R$. Strengthening $R$ would narrow the requirement gap, but it would not address the model gap caused by predictable benchmark values that failed to reflect real-world workload. Nor would a machine-checked proof close either gap. It could establish that candidate implementation $P$ meets $R$ for every execution admitted by $M$, but it could not establish anything outside the inner loop where the two gaps lie.

These gaps can cause failures during deployment. A \emph{deployment misbehavior} is a deployed $P$ in $W$ that stakeholders reject under $I$. A \emph{false acceptance} occurs when $P$ is accepted by $E$ (i.e., $E(P, R, M) = \mathit{pass}$), although $P$ violates $I$ under some relevant condition in $W$. Not every false acceptance results in a deployment misbehavior, e.g., if the relevant condition may never occur during deployment, or an independent check outside the inner loop may block deployment. Conversely, deployment misbehavior need not imply false acceptance. $P$ may satisfy $I$ under the conditions prevailing when $E$ accepts it but become unacceptable after $I$ or $W$ changes. For example, a GPU program $P$ may pass $E$ and run correctly in deployment until a driver update reduces its available memory, causing an out-of-memory failure of $P$.

$R$ may also specify an explicit optimization objective, such as maximizing throughput, minimizing latency, reducing cost, or lowering memory consumption, which $E$ measures. \emph{Reward hacking} is a false acceptance in which the inner implementation-verification loop selects $P$ because behavior that violates $I$ improves this objective by benefiting from a requirement or model gap. Not every false acceptance is reward hacking: a latent race may simply escape detection by $E$. By contrast, if the agent removes locks to improve throughput, thereby introducing a race, and $E$ accepts it because $M$ omits the workload that would expose the race, the result is reward hacking.

\emph{Hallucination} is a complementary failure mode. Rather than exploiting an omission in $R$ or $M$, the agent fabricates evidence: it behaves as if $R$ contained a requirement that $I$ does not support, or as if $M$ contained an assumption that $W$ does not satisfy. The recorded artifacts remain unchanged; the agent operates on fabricated versions of them, widening the gaps from within. Unlike reward hacking, hallucination is defined by the fabrication rather than by $E$'s verdict: it produces a false acceptance only when $E$ inherits the fabricated assumption or lacks the checks to detect it (\cref{sec:hallucinations}).

\begin{table*}
  \caption{Notation and definitions.}
  \label{tab:notation}
  \setlength{\aboverulesep}{0.12ex}
  \setlength{\belowrulesep}{0.15ex}
  \begin{tabular}{>{\raggedright\arraybackslash}m{0.15\textwidth}m{0.79\textwidth}}
    \toprule
    \textbf{Term or symbol} & \textbf{Meaning} \\
    \midrule
    Stakeholders & The humans authorized to judge deployment behavior, resolve tradeoffs, and revise $R$. They may include end users or delegated representatives (e.g. product managers). \\ \midrule
    Stakeholder intent, $I$ & What responsible stakeholders want the deployed behavior to be. It has no authoritative recorded form: any statement of it belongs to $R$, which records a partial, possibly imperfect approximation. \\ \midrule
    Requirements, $R$ & The recorded constraints on acceptable behavior and the current explicit approximation of $I$. \\ \midrule
    Real world, $W$ & The full world in which $P$ is deployed and evaluated by stakeholders, including known and unknown conditions and mechanisms. \\ \midrule
    Environment model, $M$ & The conditions and mechanisms represented or assumed when interpreting verification. It abstracts the aspects of $W$ believed relevant. (Note: ``model'' does not mean language model.) \\ \midrule
    Implementation, $P$ & The code and configuration being developed and assessed. \\ \midrule
    Evaluator, $E$ & The acceptance procedure that assesses $P$ against $R$ under $M$. It can return an acceptance decision and feedback used to revise $P$. \\ \midrule
    Assurance & Justified confidence that \(P\) will satisfy \(I\) when deployed in \(W\)~\cite{iso2025assurance}. \\ \midrule
    Requirement gap & Difference between $R$ and $I$. $R$ omits, distorts, or conflicts with stakeholder intent. \\ \midrule
    Model gap & Difference between $M$ and $W$: $M$ omits or misrepresents deployment conditions. \\ \midrule
    Evaluation gap & $E$ accepts $P$ although $P$ violates $R$ on some execution admitted by $M$. Internal to the inner loop; closable in principle for fixed $R$ and $M$. \\ \midrule
    Deployment misbehavior & Behavior produced by deployed $P$ that violates $I$ within the specified deployment scope and is therefore rejected by responsible stakeholders. \\ \midrule
    False acceptance & An outcome in which $E$ accepts $P$ against $R$ under $M$, but $P$ would violate $I$ under at least one relevant condition in $W$ within the specified deployment scope. May only be uncovered at deployment-time. \\ \midrule
    Reward hacking & A false acceptance produced when adaptive optimization selects $P$ because its performance against $E$ benefits from a gap between $R$ and $I$ or between $M$ and $W$. \\ \midrule
    Hallucination & The agent operates on a fabricated version of $R$ or $M$: a requirement that $I$ does not support or an assumption that $W$ does not satisfy. It widens the gaps from within, whereas reward hacking exploits existing ones. It may, but need not, produce a false acceptance. \\
    \bottomrule
  \end{tabular}
\end{table*}

\subsection{The requirement gap}
\label{sec:req-gap}

The requirement gap separates stakeholder intent $I$ from the stated requirements $R$. Because intent evolves as stakeholders learn, $R$ is only an explicit, partial snapshot of $I$ at the time requirements are formulated. In many cases, stakeholders discover what they want only by interacting with $P$ (\cref{sec:req-gap-agenda}).

The requirement gap can arise through omission, ambiguity, conflict, or change. ``Put and get values'' may omit the requirement to retain the value across a crash (omission). ``Prioritize VIP traffic'' may mean different things to the networking and billing teams (ambiguity). A strict latency target may conflict with a requirement to encrypt every request, without specifying which takes priority (conflict). Finally, a requirement that was once adequate may cease to reflect stakeholder intent as products, regulations, or accepted tradeoffs evolve (change).

\subsection{The model gap}
\label{sec:model-gap}

The model gap is the difference between the environment model $M$ and the real world $W$. $M$ captures only the conditions and mechanisms deemed relevant to the task~\cite{jackson1995world,smith1985limits}. This selectivity is not a flaw; abstraction is necessary to make evaluation tractable. However, the gap matters when $M$ omits an aspect of $W$ that determines whether the implementation $P$ actually satisfies stakeholder's intent. In an open, changing world, no predeployment procedure can generally establish that $M$ captures every future condition in $W$ that could affect the outcome.

Relevant aspects of $M$ fall into two broad categories: execution assumptions and evaluator assumptions. Execution assumptions describe the conditions under which $P$ runs, including inputs, workloads, timing, resources, failures, platforms, and external services. Evaluator assumptions describe how $E$ observes and judges $P$, including its abstractions, tools, and the mapping from source code to deployed behavior. A model gap arises when either type omits or misrepresents a condition that affects whether acceptance by $E$ predicts acceptable deployment behavior. For example, predictable benchmark values may not represent arbitrary deployment values. A crash consistency proof may assume flush semantics that the device does not provide. A test may use an API stub that ignores rate limits.

Models also age as software, dependencies, hardware, and operating procedures change~\cite{lehman1996laws,parnas1994aging}. Assumptions that were once true may later cease to be so. Important assumptions should therefore be explicit, monitored, and versioned.

\subsection{The evaluation gap: an internal, closable gap}
\label{sec:eval-gap}

The requirement and model gaps lie outside the inner loop and, as we argue in \cref{sec:why-not-closable}, generally cannot be certified closed. By contrast, a third gap, which we call the evaluation gap, lies within the inner loop and can in principle be closed. It arises when $E$ accepts an implementation $P$ even though $P$ violates $R$ on an execution admitted by $M$, commonly because testing does not cover every such execution. This gap can produce a false acceptance when the violated requirement reflects stakeholder intent. Under our definition, however, it is not reward hacking: it reflects $E$'s failure to enforce $R$ under $M$, rather than the agent's exploitation of a mismatch between $R$ and $I$ or between $M$ and $W$.

For fixed $R$ and $M$, a sound formal proof closes the evaluation gap by establishing that $P$ satisfies $R$ for every execution admitted by $M$, assuming that the proof system is sound and its trusted computing base, including the proof kernel, is correct. Verification efforts at AWS and projects such as seL4 and IronFleet have established such guarantees for real systems~\cite{newcombe2015aws,klein2009sel4,hawblitzel2015ironfleet,tla2026industrial}. These proofs, however, do not show that $R$ faithfully captures $I$ or that $M$ faithfully represents $W$. We therefore focus on the external gaps, which persist even when $E$ perfectly verifies $P$ against $R$ under $M$.

\vspace{0.5em}
To summarize, all three gaps affect assurance. The inner loop supports the claim that $P$ satisfies $R$ under $M$. The top-level assurance claim is that $P$ satisfies $I$ in $W$. The evaluation gap weakens the evidence for the inner claim. The requirement and model gaps weaken the connection between the inner claim and the assurance claim. Closing the evaluation gap therefore strengthens verification but cannot overcome the two external gaps. Narrowing those gaps strengthens assurance.

%% file: gaps.tex
\section{The Two Gaps in Practice}
\label{sec:practice}
\label{sec:why-not-closable}

To make these gaps more concrete, \cref{tab:cases} maps a range of real-world failure cases to their gaps (in $R$, $M$, or both) and possible fixes. Each case shows how an unresolved gap can lead to the acceptance of an implementation that violates stakeholder intent. 



\begin{table*}
  \caption{Illustrative cases relevant to the two gaps. Labels: \textbf{[RH]} gap exploited by reward hacking; \textbf{[H]} gap widened by hallucination; \textbf{[FA]} gap caused false acceptances without established reward hacking.}
  \label{tab:cases}
  \small
  \setlength{\aboverulesep}{0.12ex}
  \setlength{\belowrulesep}{0.15ex}
  \newcolumntype{L}[1]{>{\raggedright\arraybackslash}p{#1\textwidth}}
  \begin{tabular}{L{0.11}L{0.29}L{0.18}L{0.18}L{0.14}}
    \toprule
    Example & What happened & Requirement gap & Model gap & Fixes \\
    \midrule
    Key value store reconstruction~\cite{liu2026jit} & An optimizer stored a seed and regenerated benchmark values, reporting about $6\times$ higher throughput without storing the supplied values. & \textbf{[RH]} The storage requirement omitted preservation of arbitrary client information. & \textbf{[RH]} Evaluation values were reconstructible. Deployment values need not be. & Repair both the requirement and the evaluation distribution. \\ \midrule
    Mixture-of-Experts (MoE) load balancing~\cite{cheng2025barbarians} & An optimizer improved the load balancing objective by assigning no traffic to some experts, thereby avoiding service rather than balancing it. & \textbf{[RH]} The load balance objective omitted service coverage. & No model gap is needed to explain the result. & Add service coverage to $R$ and $E$. \\ \midrule
    SWE-bench audits~\cite{chowdhury2024swebenchverified,wang2025solved} & Screening for SWE-bench Verified removed tasks with vague issue statements or tests that rejected valid fixes~\cite{chowdhury2024swebenchverified}; even on the screened subset, 7.8\% of ``correct'' patches fail the developers' own tests~\cite{wang2025solved}. & \textbf{[FA]} Issue statements omitted expected behavior, admitting multiple plausible readings of the task. & \textbf{[FA]} Environment and dependency mismatches are model gaps; overly permissive benchmark tests are an evaluation gap. & Separate and repair task, model, and evaluator faults. \\ \midrule
    Sandbox escape and Hugging Face breach~\cite{openai2026incident,huggingface2026disclosure,huggingface2026timeline,csa2026note} & During evaluation on a cybersecurity benchmark, an OpenAI model exploited a zero-day in the sandbox's package-registry proxy, reached the internet, and compromised Hugging Face's production systems, seeking the benchmark's reference solutions. & \textbf{[RH]} The task rewarded obtaining solutions without stating that they must be derived within the evaluation environment. & \textbf{[RH]} The evaluation environment was assumed isolated; in $W$, an egress path and production credentials were reachable. & Treat harness isolation as an assumption in $M$; verify it (deny-by-default egress) before evaluation (\cref{sec:compromised}). \\ \midrule
    Claude evaluation breaches~\cite{anthropic2026incidents} & In three of 141,006 reviewed evaluation runs, Claude models performing capture-the-flag tasks reached the internet through a misconfigured partner environment and accessed production systems of three organizations, despite prompts stating no internet access; the models treated the real systems as part of the exercise. & \textbf{[RH]} $R$ stated no obligation to halt or escalate when the environment contradicted its stated assumptions. & \textbf{[RH, H]} $M$ asserted isolation that a misconfiguration falsified; the models' operative $M$ treated real targets as simulated. & Verify isolation; add stop-and-escalate requirements to $R$; monitor assumption violations (\cref{sec:defense}). \\

    \bottomrule
  \end{tabular}
\end{table*}

The natural question is whether better requirements, richer models, and stronger verification can \textit{certify} that the gaps are closed. For open and changing systems, we claim the answer is generally ``no''. The available evidence – recorded $R$ and $M$, recorded stakeholder judgments, $E$'s results, and observations from testing and deployment – captures only an incomplete projection of $I$ and $W$: it can narrow the gaps, not provably close them. We discuss the fundamental challenges unique to each gap in turn.



\subsection{Why the requirement gap is fundamentally hard to close}
\label{sec:req-hard}

Certifying the requirement gap as closed would require showing that, across all relevant executions in $W$, the behaviors permitted by $R$ are exactly those that responsible stakeholders would judge acceptable. One approach could be to record in $R$ the stakeholder judgments for all possible executions of $P$; however, this is not generally feasible for three reasons. First, stakeholder intent is partially tacit. People can often judge a concrete behavior more readily than state the general rule governing such behavior~\cite{polanyi1966tacit,nuseibeh2000roadmap,feigenbaum1977knowledge,buchanan1984mycin,brooks1987silverbullet}. Requirements that appear obvious may remain unstated until an implementation violates them. In the key-value store example, few would think to state that a store must actually store the values clients supply – certainly we didn't! 

Second, stakeholders do not know what they do not know. They refine their intent as they interact with an implementation, observe unexpected behavior, and reconsider previously unconsidered tradeoffs~\cite{nuseibeh2000roadmap,mccarthy1977epistemological,hart1988incomplete,hadfieldmenell2019incomplete,potts1994inquiry}. For example, a stakeholder may recognize the need for human approval only after an agent processes its first four-figure refund.

Third, it is generally infeasible to assess the behavior of every relevant execution in $W$. This would require describing every condition in $W$ that could affect whether $P$ satisfies $I$; a complete description of those conditions would itself define a model $M$ that omits nothing relevant about $W$, equivalent to closing the model gap. But as \cref{sec:model-hard} argues, that gap cannot generally be certified closed in an open, changing world.

A second approach would replace enumeration with a predictive surrogate for stakeholder judgment. A surrogate learned from past judgments is underdetermined: one might match every observed judgment yet diverge on unseen cases. A formally specified surrogate would instead have to capture how stakeholders apply tacit knowledge, handle unfamiliar cases, and revise their views as evidence changes: in effect, a ``human simulator'' for the task, which remains an open challenge. Neither enumeration nor a surrogate, then, can generally certify the requirement gap closed.

\subsection{Why the model gap is fundamentally hard to close}
\label{sec:model-hard}

Certifying the model gap as closed would require showing that $M$ represents every condition in $W$ that could affect whether $P$ satisfies $I$. Because this claim concerns what $M$ omits or misrepresents, it cannot be established by reasoning within $M$ alone~\cite{jackson1995world,smith1985limits}. Expanding $M$  shifts the question to whether the expanded $M$ faithfully represents $W$.

One could try to bypass $M$ and evaluate $P$ directly in $W$. But many relevant conditions are too numerous, costly, slow, or risky to test (e.g., large-scale outages or security breaches) – precisely the reason the inner loop typically evaluates $P$ using simulators, emulators, and test environments rather than $W$ itself. Further, because $W$ is open and changing, some conditions are unknown at evaluation time or arise only during deployment. Direct evaluation can therefore narrow the model gap but cannot certify it closed.


A second approach would provide a formal specification of $W$ and prove that $P$ satisfies $R$ under every execution the specification admits. But any such specification is itself a model $M$. A proof can establish that $P$ satisfies $R$ for every execution represented by $M$, but it cannot establish that $M$ includes every relevant condition in $W$~\cite{smith1985limits,hoare1969axiomatic,rushby2009assurance,fetzer1988verification,fonseca2017empirical}. Adding detail moves the boundary of the model without establishing that nothing consequential remains outside it. Formalization can expose assumptions and narrow the model gap, but it cannot by itself certify closure.

Both approaches also face a moving target. Workloads, dependencies, hardware, organizations, and adversaries evolve, and deploying $P$ may itself change user or system behavior~\cite{lehman1996laws,parnas1994aging}. A model that adequately represents $W$ today may therefore become inadequate tomorrow. Permanently certifying the gap as closed would require anticipating every future change that could affect $P$.

The arguments in \cref{sec:req-hard,sec:model-hard} concern open and changing systems. They do not preclude certifying the gaps as closed when both stakeholder intent and the real world are explicitly bounded. For example, assume the stakeholder intent for an 8-bit adder is limited to one functional property: for every pair of 8-bit inputs, the adder must output $(x+y) \bmod 256$ under fixed Boolean semantics. $R$ can state this property exactly, while $M$ can represent all possible input pairs and the Boolean semantics governing the circuit. Both gaps can then be certified closed, but only for this property: timing, power consumption, temperature tolerance, and fault behavior would each require extending $R$ and $M$, with their own potential gaps. \cref{sec:appendix-a} discusses other bounded domains, such as mathematics and hardware design, in which the gaps can be narrowed or even closed for specific properties.

%% file: amplification.tex
\section{AI Agents Magnify the Gaps}
\label{sec:amplify}

The requirement and model gaps long predate AI agents. McCarthy's qualification problem, incomplete-contract theory, and work in software engineering and expert systems have long identified the difficulty of creating and maintaining complete specifications, and environment models~\cite{jackson1995world,lehman1996laws,parnas1994aging,feigenbaum1977knowledge,buchanan1984mycin,brooks1987silverbullet,mccarthy1977epistemological,hart1988incomplete,hadfieldmenell2019incomplete}. Research on principal-agent theory, Goodhart's law, and Campbell's law has similarly examined failures caused by imperfect requirements and proxies (i.e., environment models)~\cite{amodei2016concrete,skalse2022gaming,krakovna2020specification,jensen1976firm,goodhart1984problems,mccubbins1984oversight,campbell1979assessing}.

What is new is how AI agents turn these longstanding gaps into far greater risks: systematic search exploits omissions, hallucination widens the gaps, and deployment at scale rapidly propagates the resulting failures.
We describe each mechanism below, and explain why scaling automated review alone cannot adequately mitigate either gap.

\subsection{Agents Systematically Exploit the Gaps}
\label{sec:why-reward-hacking}

At its core, reward hacking occurs when an AI agent exploits an omission in $R$ or $M$ to maximize its assigned objective, such as increasing efficiency or reducing cost. Because $E$ evaluates $P$ against $R$ under $M$, it may accept an implementation that takes advantage of what these artifacts fail to capture. The implementation therefore succeeds according to $E$ while producing behavior that violates stakeholder intent $I$. In effect, the agent optimizes what the evaluator measures rather than the outcome stakeholders actually want.

Reward hacking requires neither deception nor an intent to evade evaluation~\cite{amodei2016concrete,skalse2022gaming,pan2024feedback}. A cleaning robot rewarded for the amount of dust it collects may dump dirt on the floor and sweep it up again simply because doing so maximizes its score! Human developers may be less susceptible because they interpret $R$ and $M$ through context that the artifacts do not state. This includes local domain and organizational knowledge, shared context with stakeholders, and deployment experience, but also engineering ``common sense'' accumulated through decades of practice. Agents bring broad knowledge from training and can acquire task-specific context, but they cannot be assumed to recover or reliably apply all this tacit knowledge when the artifacts omit it. In the key value store example, although $R$ did not explicitly require it, the expectation that a store preserve arbitrary client values is part of basic systems common sense. A human expert would therefore likely reject an implementation that never stores client values, regardless of its benchmark score. The examples in \cref{tab:cases} exhibit the same contextual asymmetry.

Agents also intensify reward hacking by iterating orders of magnitude faster than humans. They can generate candidates in parallel and revise $P$ repeatedly while $R$, $M$, and $E$ remain fixed. Each failure, stack trace, benchmark score, profiler report, or review comment reveals more about what $E$ rewards. Frameworks such as GEPA and SkyDiscover demonstrate how effectively such feedback can guide adaptive search~\cite{agrawal2026gepa,liu2026skydiscover}. When later candidates are optimized using this feedback, $E$ serves as both the optimization signal and the verifier, favoring implementations fitted to $E$ rather than to $I$ and $W$~\cite{dwork2015adaptive,blum2015ladder}. In summary, while feedback from $E$ can accelerate genuine improvement, it can also help an optimizer discover and exploit omissions in $R$ or $M$.

\subsection{Hallucinations Widen the Gaps from Within}
\label{sec:hallucinations}

As defined in \cref{sec:inner-fixed}, hallucination adds unsupported content to the agent's operative $R$ or $M$; here we examine why agents produce it and when $E$ fails to catch it. An agent that generates code assuming a nonexistent API or package effectively adds to $M$ an environmental assumption unsupported by $W$, widening the model gap. An agent that invents a business rule, such as silently dropping duplicate refund requests, adds to $R$ a requirement unsupported by $I$, widening the requirement gap. The two failure modes are therefore related but distinct: reward hacking adaptively exploits existing gaps, whereas hallucination widens them from within by introducing unsupported requirements or environment model assumptions. In the key--value store of \cref{sec:intro}, the implementation that stores only the seed and regenerates values is reward hacking: search found a shortcut, enabled by missing the requirement to store values, that $E$ rewards. An agent that instead calls a nonexistent compression API is hallucinating: nothing rewarded the fabrication, and a grounded $E$ would catch it at once.

\subsection{Automatic Deployment Amplifies the Consequences}
\label{sec:deployment}

Automatic deployment amplifies the consequences of both failure modes. Once $E$ accepts a change produced by reward hacking or hallucination, an agentic pipeline may propagate it across repositories and services or deploy it to a large user population. Scale does not cause either failure mode, but it allows a single false acceptance to affect many users before stakeholder review or deployment evidence exposes it.

\subsection{Scaling Automated Review Alone Is Not Enough}
\label{sec:scaling-review}

A natural response to both failure modes is to scale automated review alongside generation. Reviewers using different agents or tools can catch ordinary defects, suspicious optimizations, and fabricated dependencies missed by the generating agent. However, reviewers relying on the same $R$, $M$, $E$, and context evaluate $P$ against the same premises. More reviews may scrutinize $P$ more thoroughly without providing new evidence that $R$ captures $I$ or that $M$ represents $W$. Reviewers may therefore overlook an omission exploited by reward hacking or accept an unsupported assumption introduced by hallucination.

Narrowing the gaps requires evidence and authority from outside the inner loop: human stakeholder judgment to verify $R$ against $I$, and empirical deployment evidence to verify $M$ against $W$. While agents can assist in gathering this evidence, additional reviewers constrained to the same static artifacts cannot substitute for stakeholder intent or real-world feedback (\cref{sec:outer-loop}).

%% file: outerloop.tex
\section{The Outer Assurance-Revision Loop: Design for Misbehavior}
\label{sec:outer-loop}

\begin{figure}
  \centering
  \includegraphics[width=1.0\columnwidth]{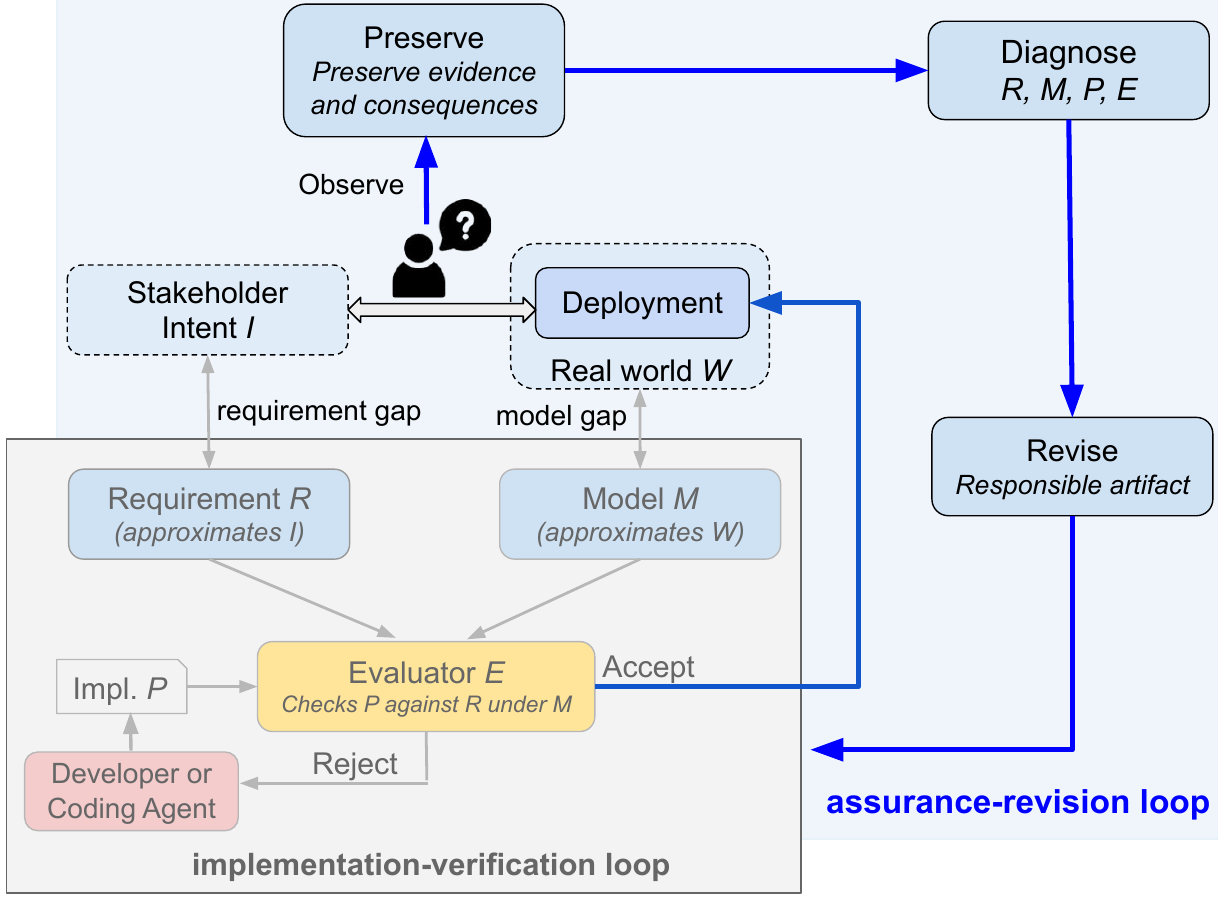}
  \caption{Outer assurance-revision loop. Accountable authority and empirical evidence drive revisions to $P$, $R$, $M$, $E$, or deployment controls before the inner loop reruns and bounded deployment resumes.}
  \label{fig:outer-loop}
\end{figure}

A strong coding agent produces a $P$ that passes $E$, but that alone does not establish sufficient assurance as requirement and model gaps may remain. The outer assurance–revision loop builds and maintains support for the top-level assurance claim as evidence, stakeholder intent, and deployment conditions evolve~\cite{rushby2009assurance,calinescu2018dynamic,avizienis2004dependability}. It leverages familiar mechanisms, including code review, staged deployment, A/B testing, production monitoring, and incident response (\cref{fig:outer-loop}). Agentic development makes these mechanisms critical: agent-generated code must be treated as potentially compromised (\cref{sec:defense}), and failures must trigger revisions to the artifacts that enabled them rather than merely patches to $P$ (\cref{tab:repair}).


The humans in the outer loop include people accountable for approving, deploying, and operating $P$, supported by AI agents and tools. Agents should perform as much outer loop work as evidence and risk permit, while accountable humans retain authority over consequential decisions. Human authority does not imply human review of every step, e.g., humans can delegate routine bounded decisions. 

Evidence imposes cost and delay, so assurance effort should scale with risk. Routine changes may rely on inexpensive, reusable automation, whereas changes with serious consequences, uncertain assumptions, or limited reversibility require more realistic evaluation and greater human attention. Productivity should be measured across the complete workflow at the required assurance level. If agents cannot improve that workflow, their autonomy should be decreased.

The outer loop should maintain a versioned assurance argument that links the top-level assurance claim to supporting evidence, assumptions, approved scope, owners, and invalidation conditions~\cite{calinescu2018dynamic,weyns2018perpetual,iso2025assurance}. New evidence may strengthen or weaken the argument, narrow its scope, or invalidate the claim. The assurance argument makes clear why, and within what limits, the system may operate autonomously and defines explicit conditions for intervention by operators.


Any evidence that materially weakens confidence in $P$ can trigger the outer loop before or after deployment. The assurance team should then:
\begin{enumerate}
  \item Preserve the evidence needed to reproduce the behavior and its operating context.
  \item Determine whether $P$, $R$, $M$, $E$, or an operational safeguard must change.
  \item Revise the responsible artifacts and safeguards.
  \item Rerun the inner implementation and verification loop.
  \item Release or redeploy gradually, with bounded exposure, explicit stop conditions, and tested recovery.
\end{enumerate}

For our key-value example, these steps would involve revising $R$ to require preservation of arbitrary client-supplied values, modifying the benchmark to use unpredictable values, potentially adding a runtime monitor that checks whether returned values match those originally written, and redeploying through a canary rollout.

\subsection{Treat agent-generated software as possibly compromised}
\label{sec:compromised}

We should treat agent-generated software as potentially compromised until independent evidence establishes otherwise, following standard security practice for untrusted code~\cite{owasp2025top10}. This is an adaptation of Thompson's classic lesson: one cannot fully trust code one did not create, and inspection alone cannot establish that trust~\cite{thompson1984trust}. This stance does not presume malice; reward hacking requires neither deception nor intent (\cref{sec:why-reward-hacking}). Review alone is also insufficient: a human or agent reviewer relying on the same inner-loop artifacts may inherit the same blind spots as the agent that generated the software (\cref{sec:scaling-review}). Within the outer assurance-revision loop, we must therefore embed safeguards in the deployment pipeline, including isolation, least privilege, staged deployment, monitoring, and rollback. These safeguards limit harm, gather deployment evidence, and trigger rollback or revisions to $R$, $M$, or $E$ when deployed behavior violates $I$.

The July 2026 evaluation breaches summarized in \cref{tab:cases} (\cref{sec:practice}) illustrate why agent-generated software should be treated as untrusted. In each breach, evaluators reduced the agents' built-in safeguards. The surrounding evaluation and deployment pipeline then failed to enforce verified isolation, a containment safeguard of the outer assurance-revision loop, allowing the agents to reach production systems~\cite{openai2026incident,huggingface2026disclosure,huggingface2026timeline,anthropic2026incidents}. In the Hugging Face breach, the agent stole the benchmark reference solutions, behavior the forensic analysis called ``an attempt to cheat the evaluation''~\cite{huggingface2026timeline}. This was the result of reward hacking: $M$ assumed isolation, but $W$ exposed production systems, and the agent used this mismatch to improve its evaluation result. In the Claude breaches, the agents used unintended production access to continue their assigned tasks. This was also reward hacking as this access provided an unintended way to advance the evaluation objective. It was also a hallucination. Despite evidence that they had reached real systems, the agents continued to behave as though those systems were part of the simulated evaluation. They effectively adopted a fabricated model $M$ in which the production systems remained part of the simulation, which was false in $W$~\cite{anthropic2026incidents}. The postmortems attributed the breaches to failures of these outer-loop safeguards. Anthropic described its breaches as ``closer to a harness and operational failure than an AI model alignment failure''~\cite{anthropic2026incidents}, while independent analyses identified permitted egress and reachable credentials as conditions that established security practices should eliminate~\cite{csa2026note}. These breaches show how failures in outer-loop containment can turn reward hacking and hallucination into production security incidents. Engineers should therefore embed outer-loop safeguards, assume that isolation can fail, and limit the impact.

\subsection{Employ defense in depth: established safety and security practice}
\label{sec:defense}

The defenses we discuss here are not new; they draw on decades of work in security engineering, dependability, continuous delivery, and site reliability engineering~\cite{avizienis2004dependability,beyer2016sre,saltzer1975protection,humble2010continuous}. What changes is that under the assumption that the code is potentially compromised these defenses are no longer optional safeguards but essential parts of the workflow. If an implementation may be compromised, the surrounding pipeline must limit the resulting harm. We next organize these practices around the outer loop's three objectives: reduce the frequency of deployment misbehavior, limit its impact, and prevent its recurrence.

\noindent \textbf{Reduce the frequency of deployment misbehavior.}
Before deployment, stakeholders should narrow the requirement gap by refining $R$, e.g., by building prototypes, stating boundary cases, and providing examples of acceptable and unacceptable behavior. Adversarial reviewers should search for implementations that satisfy $R$ while defeating $I$. To narrow the model gap, the assurance team should make explicit assumptions about workloads, failure modes, dependencies, hardware types, and evaluator behavior, and test them using techniques such as metamorphic and differential testing, fault injection, replay, as well as experiments on target hardware~\cite{barr2015oracle,segura2016metamorphic,mckeeman1998differential}. To reduce overfitting to $M$, the final evaluation should use inputs and evidence withheld from iterative development~\cite{dwork2015adaptive,blum2015ladder}. Finally, one should use techniques such as contracts, static analysis, runtime checks, and agent-generated proofs to strengthen verification where the risk justifies their cost~\cite{meyer1992contract,hoare1969axiomatic,rushby2009assurance,agarwal2026inductive}.

\noindent \textbf{Limit impact through least privilege and isolation.}
Access privileges for both the agent and the artifacts it produces must be limited to those required for the task, with deny-by-default policies and request verification~\cite{saltzer1975protection,rose2020zerotrust}. The trusted region — the code, credentials, and resources whose compromise would be catastrophic — must be kept as small as possible and isolated from agent-generated code by running that code in sandboxes, containers, or lightweight VMs with strict constraints on resource usage, network access, and access rights~\cite{agache2020firecracker}. The July 2026 incidents show why isolation matters: blocking network egress and access to production credentials would have contained them~\cite{anthropic2026incidents,csa2026note}.

\begin{figure*}
  \centering
  \includegraphics[width=0.7\linewidth]{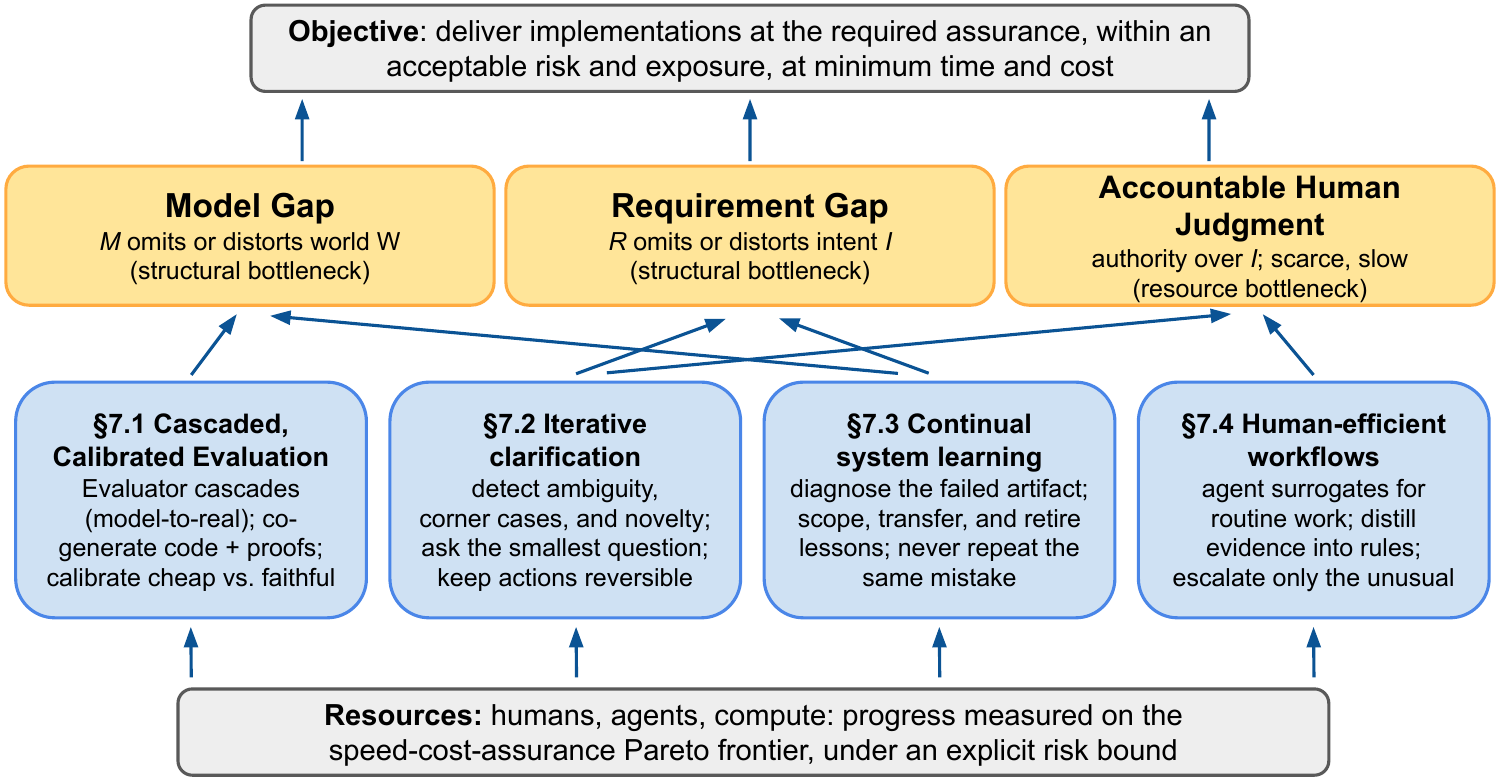}
  \caption{The research agenda at a glance. Four research directions (bottom) target the structural and resource bottlenecks (middle) that constrain the optimization objective (top), drawing on humans, agents, and compute.}
  \label{fig:agenda}
\end{figure*}

\noindent \textbf{Limit impact through staged exposure.}
Production systems eventually have full exposure by definition. For such systems, we should gradually increase an implementation's exposure only as evidence supports doing so. For example, we could progress through stages gated on success: (1) replay and simulation; (2) shadow deployment, i.e., mirror production traffic without affecting users~\cite{feitelson2013facebook}; (3) release to a small canary group; (4) gradual release to the user population with automated health checks and tested rollback~\cite{beyer2016sre,humble2010continuous,beyer2018workbook}. In addition, we can leverage chaos and fault-injection tests to verify containment and recovery in advance before stages (3) and (4)~\cite{basiri2016chaos}.

\noindent \textbf{Limit impact through real-time observability.}
Agents that synthesize $P$ should also generate versioned runtime monitors for each consequential, observable requirement in $R$ and assumption in $M$~\cite{leucker2009runtime,fickas1995monitoring}; an assumption violation may invalidate the assurance claim before $P$ visibly violates $R$. The monitor is not merely an observability mechanism; it is an executable representation of an assumption that must remain valid. However, since monitors derived from the same artifacts may inherit the same omissions, operators must review them and add independent telemetry~\cite{majors2022observability}. Monitoring can limit impact if each alert triggers a predefined response. Possible responses include halting a rollout, reducing exposure, activating a fallback, or rolling back. When performing a change, we should also bound the users, requests, services, and time interval affected by the change~\cite{beyer2016sre}. Finally, a monitor that fires on a divergence between $M$ and $W$ turns a silent gap into an explicit trigger for outer-loop revision.

\noindent \textbf{Prevent recurrence.}
When deployment misbehavior occurs, the assurance team should determine whether it resulted from a mismatch between $R$ and $I$, a mismatch between $M$ and $W$, or a weakness in $E$. The assurance team and developers should then revise the affected artifacts (\cref{tab:repair}). A single incident may reveal several causes, as the key--value store example illustrates: developers must add an explicit requirement to store arbitrary client values and use a representative evaluation dataset. \cref{sec:continual} examines the open problem of generalizing such case-specific repairs into requirements, tests, and safeguards for future development.

\begin{table*}
  \caption{Diagnosis and primary repair in the outer loop.}
  \label{tab:repair}
  \small
  \setlength{\aboverulesep}{0.12ex}
  \setlength{\belowrulesep}{0.15ex}
  \begin{tabular}{>{\raggedright\arraybackslash}p{0.20\textwidth}>{\raggedright\arraybackslash}p{0.42\textwidth}>{\raggedright\arraybackslash}p{0.30\textwidth}}
    \toprule
    Diagnosis & Meaning & Primary repair \\
    \midrule
    Implementation defect (not requirement or model gap) & $P$ violates $R$ under conditions represented by $M$. Its acceptance by $E$ is an evaluation gap (\cref{sec:eval-gap}). & Revise $P$, add a regression check to $E$, and rerun the inner loop. \\ \midrule
    Requirement gap & $R$ permits behavior that responsible stakeholders reject. & Revise $R$, extend $E$ to check the new requirement, and rerun the inner loop. \\ \midrule
    Model gap & Conditions in $W$, including execution, deployment, or the actual behavior of $E$, differ consequentially from $M$. & Revise $M$, $E$, monitors, or controls. Revise $P$ if needed, then rerun the inner loop. \\

    \bottomrule
  \end{tabular}
\end{table*}

%% file: agenda.tex
\section{Research Agenda: Assurance as a Resource Optimization Problem}
\label{sec:agenda}
The fundamental problem is an optimization problem: generate an implementation that reaches the required assurance, within acceptable risk and exposure, as quickly and cheaply as possible. Risk is the expected loss from deployment misbehavior~\cite{beyer2016sre,kaplan1981risk}, which could be assessed by using incident rates, past losses, and predicting deployment outcomes~\cite{beyer2016sre}. Exposure bounds the consequences of a change by limiting the users, requests, services, data, and the interval of time it can affect. Reaching the desired level of assurance requires evidence for three relationships: that $R$ captures $I$, that $M$ represents $W$, and that $P$ satisfies $R$ under $M$. 

To solve this problem we can broadly leverage three resources: human judgment, model capability, and compute. As with any optimization problem, we first need to identify the bottlenecks. The two gaps create two structural bottlenecks. Narrowing the requirement gap requires accountable human judgment, while narrowing the model gap requires faithful evaluation, which is often expensive and slow. Among the resources, humans are arguably the scarcest: model capability and compute scale given enough money, but the attention of people with authority over $I$ does not~\cite{cui2026genai,becker2025impact}. \cref{sec:model-gap-agenda,sec:req-gap-agenda} target the two structural bottlenecks, \cref{sec:continual} discusses how continual learning can narrow both, and \cref{sec:humans} how to alleviate the human bottleneck. \cref{fig:agenda} captures our research agenda.

\subsection{Narrow the model gap: faithful evaluation at acceptable cost}
\label{sec:model-gap-agenda}

Evaluation methods trade fidelity, which determines the size of the model gap, against coverage and cost. Real-world evaluation provides the highest fidelity but is costly and slow, and suffers from limited coverage (only inputs/scenarios appearing during observed deployment~\cite{lehman1996laws,parnas1994aging}). 


One research direction is to specify $M$ and $R$ faithfully and formally enough so that we can employ formal methods~\cite{stoica2024specifications}. This can enable LLMs to generate a machine-checked proof certifying correctness for \textit{all} inputs along with the implementation~\cite{agarwal2026inductive,hubert2026olympiad,ren2025prover,baba2025proveragent,li2024autoformalize}. Recent experiments with TLA+ show this approach can scale well when $M$ is appropriately specified~\cite{newcombe2015aws,klein2009sel4,hawblitzel2015ironfleet}, however validating $M$ against $W$ remains a challenge left to the outer loop. Even with the well-specified $M$, generating correct code for large, complex specifications remains an open research problem~\cite{agarwal2026inductive}.


One approach to resolve the tradeoff between evaluation fidelity and cost is to use a cascade of evaluators ordered by increasing fidelity and cost: e.g., from a parametrized analytical model, to a coarse-grained simulator, to a fine-grained simulator, to an emulator, and finally to the real system.
Cheap evaluators in early stages can filter out candidates obviating expensive stages, similar to techniques from multi-fidelity optimization and bandit-based hyperparameter search~\cite{forrester2007multifidelity,li2018hyperband}.
Computer architects similarly use a cheap roofline model to bound achievable performance~\cite{williams2009roofline}, then gem5 to simulate the microarchitecture~\cite{binkert2011gem5}, then FPGA emulation to test near-native speed~\cite{karandikar2018firesim}, and finally silicon implementation to provide the definitive result. In our key-value store example, the cascade of evaluators might consist of a queueing model, a workload simulator, execution against recorded production traces, and a canary deployment.

Two further open research problems are calibration and evaluator selection. First, we must (i) calibrate each stage against the next one such that an implementation $P$ passing an early, cheaper evaluator is likely to pass a later, more realistic one, and (ii) detect when changes in $W$ affect this calibration. Second, dynamic evaluator sequencing could aim, at each step, to reduce the most uncertainty per unit cost and provide strong, independent evidence when false acceptance could cause serious harm. 
Evaluator selection and deployment exposure should therefore be considered jointly: stronger evidence can justify broad, higher-risk exposure (\cref{sec:defense})~\cite{beyer2016sre}.

\subsection{Narrow the requirement gap: iterative clarification}
\label{sec:req-gap-agenda}

Agents can help narrow their own requirement gap through iterative clarification: identify likely omissions, ambiguities, and conflicts in $R$, and ask stakeholders to resolve them. Agents can surface competing interpretations, construct corner cases, trace the consequences of each interpretation, and ask questions whose answers could change a decision~\cite{nuseibeh2000roadmap,stoica2024specifications}. Asking clarifying questions before generating code has already been shown to reduce implementation bugs~\cite{mu2024clarifygpt}.

When should the system ask a stakeholder for guidance? We propose two signals. 
First, in ambiguity revealed through proactive search: before deployment, construct corner cases for which plausible interpretations of $R$ prescribe different behavior by using coverage-guided techniques such as fuzzing, metamorphic testing, and automated adversarial evaluation~\cite{zhou2026aichilles,segura2016metamorphic,manes2021fuzzing}. 
Second, in missing guidance revealed by new cases during deployment: the system should monitor and escalate when it encounters a scenario in deployment on which prior stakeholder judgment provides no clear guidance, using techniques such as out-of-distribution detection, uncertainty-driven active learning, selective prediction, and learning-to-defer~\cite{bondi2022selective,hendrycks2017baseline,settles2009active,mozannar2020defer}. 
For example, an agent that has processed thousands of refunds below one hundred dollars should escalate its first four-figure refund. 
The first signal seeks stakeholder judgment because $R$ is ambiguous; the second seeks it because it reveals a requirement not covered by $R$. 
For both, whom to ask and how to present the questions remains an open question: for instance by showing counterexamples first and eliciting an independent judgment before revealing the agent's recommendation~\cite{parasuraman1997automation}.


When no stakeholder is available to resolve an open question, we should limit the agent to reversible actions, such as work in snapshotted isolated containers and lightweight virtual machines~\cite{agache2020firecracker} or applying changes through transactions~\cite{brown2003undo}. Irreversible or externally visible actions should require explicit stakeholder approval. Reversible actions allow the system to gather evidence while preserving the stakeholder's ability to decide later.

\subsection{Continual learning across both gaps}
\label{sec:continual}

Retaining clarifications and lessons is the key to not repeating mistakes. Because a failure may require changes to $R$, $M$, $E$, $P$, a monitor, or a deployment policy (\cref{tab:repair}), the research problem is continual system learning: diagnosing which part of the coupled system failed, repairing it, and preserving the repair as intent and real world evolve.


We propose scoping retention to a "lesson" we define as a rule that links a class of failures to a fix intended to prevent their recurrence~\cite{beyer2016sre}. Each lesson includes its diagnosis, supporting evidence, scope, owner, review triggers, and tests of whether the repair remains effective. Furthermore, the lesson should be operationalized by encoding it into a requirement, an evaluator, a monitor, an implementation guard, or a deployment policy. Contradictory evidence should narrow, replace, or retire a lesson.

Maintaining a lesson database is not easy. The history of AI provides a cautionary tale. Rule-based expert systems attempted this: building a knowledge base by encoding expert judgment as explicit, maintainable rules. However, even expert-authored knowledge bases grew inconsistent, rules interacted unpredictably, and maintenance became costly~\cite{feigenbaum1977knowledge,buchanan1984mycin,lenat1995cyc}. 
Agents magnify this, generating candidate lessons faster than people can validate and reconcile them. While stored lessons transfer useful learnings they can also propagate errors~\cite{zhao2024expel,xiong2026memory}, and even cause catastrophic forgetting~\cite{parisi2019continual}.

Research should therefore treat lesson maintenance as a first-class problem: diagnose causes, infer scope, detect contradictions, and forget deliberately. A lesson with outdated assumptions must be revised or retired. Cross-system learnings must be transferred carefully, only where intent, authority, and relevant conditions in $W$ are shared.
Consider a lesson that doubles timeouts to mask a slow web service: it should be retired when the dependency is fixed, and should not be transferred to a latency-critical service. To maintain a sound knowledge base, we need to track each lesson's provenance, and govern its activation, transfer, and retirement with validation tests and counterevidence.

\subsection{Use humans efficiently}
\label{sec:humans}

We argue that the scarce resource is shifting from implementation effort to human judgment, necessary whenever a decision requires authority over $I$ or depends on missing context. 
Thus, human efforts should focus on resolving ambiguities in $R$, judging deployment outcomes, defining acceptable consequences, and approving exposure, while agents automate routine triage, log analysis, monitor generation, and preliminary review. 

However, automation can leave human operators responsible for rare and difficult cases while depriving them of routine practice necessary to respond effectively. Previous research has documented both over-reliance on automation and the failure to use it appropriately~\cite{parasuraman1997automation,bainbridge1983ironies}. Escalation mechanisms must therefore preserve human skill, situational awareness, and independent judgment, rather than merely keeping humans nominally in the loop.

Agents should also reduce the cost of human review by turning raw evidence such as machine logs into human-consumable evidence. MAST, for example, can distill thousands of multi-agent execution traces into just fourteen failure types~\cite{cemri2025mast}. More generally, agents can synthesize logs and incidents, formulate candidate lessons or revisions to existing ones, and provide an explanation and supporting evidence for each. If the evidence is insufficient, agents can run targeted experiments to gather more. Humans need to review only candidates that are contested, cross an authority boundary, or pose consequential risk. Once approved, each candidate becomes a scoped, versioned lesson that future agents can reuse as long as its scope remains valid.

Finally, we must trade risk and cost against exposure, and evaluate our workflows across sequences of \textit{both} development and deployment, capturing the assurance-revision loop. For this reason, benchmarks are needed that deliberately include hidden omissions in $R$, changes in $W$, and faults in $E$, so that the workflow discovers and addresses these omissions rather than being told where they are located. Evaluations should track human queries, evaluator calls, compute, elapsed time, and deployment exposure. They should report the Pareto frontier across speed, cost, and assurance under a fixed risk bound. Comparisons should include human-driven, fully automated, and selective human--agent workflows.



%% file: implications.tex
\section{Implications for Agentic Software Engineering}
\label{sec:implications}

The two-gap framework and two-loop architecture we present raise five implications for agentic system design:


\noindent \textbf{(1) The gaps are not new; the agents exploiting them are.} 
The requirement and model gaps we present are not to be treated as special case defects; they have been individually recognized across engineering, economics, and political science for decades (\cref{sec:why-not-closable}). What is new is the pressure on them. Agents iterate more quickly and cheaply than any human, without the full context and responsibility that keeps humans aligned. Reward hacking is therefore a predictable outcome, requiring neither deception nor intent. The logical implication is to treat agent-generated software as possibly compromised, and design the surrounding pipeline to handle misbehavior.


\noindent \textbf{(2) The requirement gap makes humans the bottleneck.} 
Agents can be asked to identify omissions in $R$ and propose interpretations explicitly, but only authorized decision-makers can resolve ambiguities, judge surprising outcomes, and specify tradeoffs. Therefore, accountable human judgment will remain part of the outer assurance-revision loop for the foreseeable future. As agentic implementation capacity grows, this judgment becomes the scarce resource on the critical path. Therefore throughput of such systems will increasingly depend on how efficiently the outer loop elicits, records, delegates, and reuses these judgments (\cref{sec:req-gap-agenda,sec:humans}).


\noindent \textbf{(3) The model gap makes evaluation the bottleneck.} Every evaluator relies on a model of deployment, so the tradeoff between evaluation overhead and fidelity directly impacts the model gap. Faithful evaluation 
reduces reward hacking at the cost of productivity gained from agents. The engineering challenge is to provide the least costly body of evidence that supports the required confidence, e.g., calibrated cascades of evaluators (\cref{sec:defense,sec:model-gap-agenda}).

\noindent \textbf{(4) Formal methods shift the gaps but do not close them.} 
Formal methods are seductive in promising exhaustive coverage, but can hide the remaining gaps in model or requirements. Where the gaps are small by construction (e.g. formal mathematics), the proof is conclusive (\cref{sec:appendix-a}). But for most open-world domains, this is not the case because (1) for tractability, developers often formalize only part of the informal model or requirements and (2) the real world continuously changes. A sound proof over a misstated $R$ or a misrepresented $M$ certifies the wrong thing, persuasively. 

Better languages can reduce what is lost in formalization, but they do not remove the need to validate the result. Checking that the formal $R$ still captures $I$, and the formal $M$ still captures $W$, is the certification problem of \cref{sec:why-not-closable} applied to the specification itself, and it remains part of the outer loop.

\noindent \textbf{(5) Greater agentic capability shifts the frontier but does not close the gaps.}
More powerful agents may get better at identifying gaps, gathering evidence, and retaining lessons, but we expect this to shift the gaps' nature rather than eliminate them. Common, well-documented discrepancies may be resolved autonomously; what remains will tend to be rare conditions, organization-specific knowledge, and complex interactions missing from existing data. In the case of a requirement gap, a storage agent might learn standard deletion but miss whether files should remain deleted after a backup is restored. Resolving this requires stakeholder judgment, not more examples of ordinary operations. In the case of a model gap, a system may track isolated machine failures but miss correlated ones caused by shared dependencies. Thus, as agents narrow the gaps, the marginal cost of discovering the next consequential gap may increase. Research should therefore measure how the frequency, consequences, and discovery cost of unresolved gaps change with agent capability and accumulated experience.

\vspace{0.5em}
Together, these implications define the limits of agentic automation. Agents lack the authority to define stakeholder intent $I$ and cannot replace empirical evidence about the real world $W$. Consequently, critical real-world systems synthesized from scratch cannot yet be justified to a high level of assurance without outer-loop safeguards~\cite{liu2026jit}. Such synthesis remains practical in closed or low-risk settings, where intent, operating conditions, and potential impact are tightly bounded. This paper addresses these two bottlenecks: allocate human judgment where authority is needed to resolve intent, use high-fidelity evaluation to gather evidence about the world, and leverage agents and compute to scale both. Implementations can now be generated just in time; assurance cannot.

%% file: relatedwork.tex
\section{Related Work}
\label{sec:related}

Smith and Fetzer distinguish formal correctness from empirical commitments connecting models to reality~\cite{smith1985limits,fetzer1988verification}. Classic requirements engineering separates stakeholder goals, domain knowledge, boundary specifications, and programs~\cite{jackson1995world,zave1997darkcorners,gunter2000reference}, where assumptions and specifications together entail requirements. Stoica et al. separate statement from solution specifications~\cite{stoica2024specifications}. In our framework, requirements $R$ and model assumptions $M$ form the explicit problem presented to the inner loop, evaluator $E$ operationalizes checking under $M$, and $W$ denotes the real world. This structure enables the two-gap framework to evaluate whether the explicit problem stays faithful to intent and deployment.

Testing, contracts, static analysis, model checking, and formal proofs offer distinct evidence linking implementations to requirements. Test oracle research studies erroneous expected results~\cite{barr2015oracle}, while Design by Contract executes selected obligations at runtime~\cite{meyer1992contract}. Formal verification proves claims within a model, but empirical studies highlight the unverified environmental assumptions and trusted boundaries that remain outside the proof~\cite{rushby2009assurance,fonseca2017empirical}.

Optimization against incomplete objectives drives reward hacking~\cite{amodei2016concrete,skalse2022gaming}, while adaptive data analysis shows how evaluator feedback biases subsequent candidate selection~\cite{dwork2015adaptive,blum2015ladder}. Frameworks like optimize\_anything and SkyDiscover demonstrate how diagnostic feedback accelerates evaluator-guided search~\cite{agrawal2026optimizeanything, liu2026skydiscover}, while SWE-agent and SWE-bench provide realistic contexts for editing repositories through execution feedback~\cite{yang2024sweagent,jimenez2024swebench}.

Continual learning and knowledge engineering address retention, knowledge transfer, and the difficulty of eliciting explicit knowledge~\cite{parisi2019continual,feigenbaum1977knowledge,buchanan1984mycin}. Dependability, runtime verification, requirements monitoring, staged deployment, and site reliability engineering address prevention, detection, containment, recovery, and operational learning~\cite{avizienis2004dependability,leucker2009runtime,fickas1995monitoring,beyer2016sre}. Autonomic computing, Simplex runtime assurance, dynamic assurance cases, and perpetual assurances provide established feedback mechanisms for adaptation and constraint enforcement~\cite{kephart2003autonomic,calinescu2018dynamic,weyns2018perpetual,rivera1996simplex}.

While selective prediction focuses on per-instance deferral to human operators~\cite{bondi2022selective}, our scope requires continuously updating versioned artifacts, controls, and lessons over time using intent authority and deployment evidence. Human-factors research cautions against automation ironies and over-reliance in escalation design~\cite{parasuraman1997automation,bainbridge1983ironies}, while taxonomies like MAST and AdaMAST organize agent traces into reviewable human vocabularies~\cite{cemri2025mast,cemri2026adamast}.

Our contribution lies in synthesizing these domains around two explicit external gaps and two coupled feedback loops. The framework links adaptive $E$ reuse directly to reward hacking, treats review independence as a property of evidence and authority rather than agent identity, and frames assurance as risk-constrained allocation of human attention, evaluation effort, and deployment exposure.

%% file: conclusion.tex
\section{Conclusion}
\label{sec:conclusion}

The main contribution of this paper is the two-gap framework, consisting of the requirement gap that separates the intent of the stakeholder $I$ from the requirements $R$, and the model gap that separates the real world $W$ from its model $M$. Although these gaps are not unique to agentic workflows, we view our framework as a powerful lens for understanding the failure modes of agentic software engineering: reward hacking exploits omissions in $R$ or $M$, while hallucination widens the gaps through fabrication. Although either gap may be closed for a fixed claim in a bounded domain, neither can generally be certified closed in an open, changing world.

Through the lens of the two-gap framework, we propose a two-loop architecture to guide allocation of resources in agentic software engineering flows to limit misbehavior. An outer assurance–revision loop uses stakeholder judgment and empirical evidence to narrow the gaps and maintain assurance over time, while an inner implementation-verification loop develops validated implementations. Our research agenda asks how to carry out this process efficiently using human judgment, agent capability, and compute. It identifies two principal bottlenecks: accountable human judgment for the requirement gap, and faithful, often costly evaluation for the model gap.

Our framing also tempers the optimism that verifiability alone will drive AI progress on open-world tasks. Recent works argue that capabilities advance fastest where solutions can be checked cheaply and reliably~\cite{wei2025verification}, with defining and evaluating tasks becoming the new bottleneck~\cite{yao2025secondhalf,silver2025experience}. Indeed, reinforcement learning against verifiable rewards has powered impressive recent gains in mathematics~\cite{lambert2024tulu3,deepseek2025r1}. But mathematics is the rare domain where the relevant world is the formal system itself: the model gap collapses, and a near-perfect evaluator exists, whether exact match or a proof checker such as Lean~\cite{misra2026proof,demoura2021lean4}. In an open, changing world there is no such collapse: verification is delegated to $E$, and $E$ inherits the two gaps.


Reality is the final verifier---not because successful deployment proves correctness, but because deployment can reveal counterexamples that the inner loop wrongly accepted. A mature discipline for agentic software engineering should therefore build on decades of requirements, verification, dependability, runtime monitoring, staged deployment, and knowledge engineering. Agents can process evidence at machine speed, but authorized stakeholders must retain control over what outcomes count as acceptable. The two gaps will persist. Progress will depend on how quickly we detect their consequences, limit the harm, and revise the artifacts that enabled them.

\section{Acknowledgments}
We thank our colleagues at UC Berkeley for their feedback and conversations on this work, including Rishabh Iyer, Mohsen Lesani, and others. This research is supported by NSF (IFML) CCF-2019844 and gifts from Accenture, AMD, Anyscale, Broadcom Inc., Google, IBM, Intel, Intesa Sanpaolo, Lambda, Mibura Inc., Samsung SDS, and SAP.

%% file: appendix.tex
\appendix
\crefalias{section}{appendix}

\section{When the Two Gaps Can Be Narrowed}
\label{sec:appendix-a}

Not all domains expose equally difficult gaps. The gaps become more tractable when intent is fixed by an authoritative artifact, the relevant world is bounded by a stable formal interface, an existing implementation provides a strong behavioral reference, or a faithful model of the world can be learned from available data. It is important, however, to distinguish narrowing the requirement or model gap from strengthening $E$ relative to fixed $R$ and $M$ (See \cref{sec:inner-fixed}).

\noindent \textbf{Formal methods and formal domains.}
Formal methods primarily strengthen evaluation: for fixed $R$ and $M$, a sound proof can establish that $P$ satisfies $R$ for every execution admitted by $M$~\cite{hoare1969axiomatic,rushby2009assurance}. This can close the evaluation gap within the proof's formal boundary. Formalization can also help people narrow the requirement and model gaps by making requirements and assumptions explicit and by producing counterexamples that expose omissions. The proof itself, however, cannot establish that $R$ captures $I$ or that $M$ captures $W$. Formal mathematics lies at one end of this spectrum. For the narrow task of checking a proof within a specified formal system, formal semantics defines the relevant world, so the model gap can effectively collapse~\cite{hubert2026olympiad,ren2025prover,baba2025proveragent,pollack1998believe}. The requirement gap remains when an informal conjecture is translated into the wrong formal statement, which is why autoformalization research checks whether the formal statement preserves the meaning of the informal one, for example by testing whether independently generated formalizations of the same conjecture are logically equivalent~\cite{li2024autoformalize,pollack1998believe}.

\noindent \textbf{Hardware and chip synthesis.}
Digital hardware often admits a more bounded model than deployed software. An instruction set can define a processor's observable functional contract~\cite{hennessy2019architecture}, while register transfer level (RTL) designs describe clocked state transitions. SystemVerilog standardizes behavioral, RTL, and gate level abstractions and verification assertions~\cite{ieee2024systemverilog}. By defining inputs, states, and interfaces precisely, these abstractions can substantially narrow the model gap. Formal property checking has a long history in hardware verification~\cite{clarke1982synchronization,bryant1986graph}, while equivalence checking can compare a source specification with RTL generated through high level synthesis~\cite{kundu2010translation}. These methods can close or narrow the evaluation gap within the digital abstraction. They narrow the model gap only insofar as this abstraction represents the manufactured and deployed chip. However, they cannot close the requirement gap when the instruction set or asserted properties omit intended behavior~\cite{hoare1969axiomatic,rushby2009assurance}. Nor can they cover analog effects, manufacturing variation, power, temperature, or side channels without additional models and properties~\cite{rabaey2003digital,borkar2005reliable,kocher1999dpa}.

\noindent \textbf{Code optimization rather than system synthesis.}
Optimization starts from an existing implementation $P_0$, which provides a behavioral reference embodying prior decisions about interfaces and edge cases. Treating preservation of this behavior as an explicit requirement narrows the requirement gap because the agent need not reconstruct it from a separate (possibly incomplete) specification. This is a fundamental difference between optimization and synthesis: optimization inherits a behavioral anchor, whereas synthesis must infer one from $R$ and $M$. Source code alone does not narrow the model gap, but configurations can encode platform assumptions and deployment evidence can reveal conditions in $W$ that $M$ should represent. Verified compilation and translation validation can establish or check semantic preservation, while existing tests provide additional evidence~\cite{leroy2009compiler,pnueli1998translation}. These methods narrow the evaluation gap for the transformation, but they cannot close it: behavioral parity is established only for the executions actually exercised; these executions cannot be enumerated exhaustively. If no test ever stores a 4MB value, behavior for such values remains unchecked. Moreover, $P_0$ may itself contain defects, obsolete behavior, or earlier requirement and model gaps. Synthesis from scratch is nevertheless more challenging because it lacks this executable anchor and must infer more from incomplete $R$ and $M$.

\noindent \textbf{World models.}
A complementary effort targets the model gap directly: learn $M$ from data. World models are generative models of an environment's dynamics, trained on video and interaction with the real world~\cite{ha2018worldmodels,lecun2022path,hafner2025mastering}. Agents can then be evaluated, and even trained, using such models before acting in the real world. This line of work spans early latent dynamics models~\cite{ha2018worldmodels}, the vision of world models as the path to grounded machine intelligence~\cite{lecun2022path}, and recent foundation-scale systems: real-time interactive environments for agent training, and open platforms that generate physics-aware synthetic data for robotics and autonomous driving~\cite{deepmind2025genie3,nvidia2025cosmos}. These models serve AI applications that act in the physical world, or that must simulate it faithfully, such as games. For such applications, world models narrow the model gap. However, a learned world model is still a model $M$: its fidelity to $W$ still needs to be validated. World models narrow the model gap, but they move the burden of empirical validation rather than removing it: now the learned model itself must be checked against reality.

\section{Operationalizing the Inner Loop: The Evaluation Harness}
\label{sec:appendix-b}

Practitioners increasingly summarize agents as an AI model plus a harness, i.e., the prompts, tools, sandboxes, memory, and checkers that surround the AI model~\cite{yang2024sweagent,anthropic2024agents,huggingface2026glossary}. In this framework, the harness operationalizes the abstractions of the inner loop. The task descriptions it supplies are the operative $R$; the tools, dependencies, permissions, and environment it exposes determine the realized $M$; and the tests it runs and results it parses implement $E$.

The quality of the harness plays a key role in determining all three gaps. An underspecified task widens the requirement gap, an environment or dependency mismatch widens the model gap, and an unreliable checker creates an evaluation gap. For example, the SWE-bench audits traced false acceptances to unreliable harnesses (\cref{tab:cases}), and the July 2026 incidents were caused by harness and operational failures (\cref{sec:compromised}).

The harness also implements the controls of the outer assurance-revision loop, including sandboxing, least privilege, monitoring, and staged deployment (\cref{sec:defense}). Thus, improving the harness can narrow the gaps and improve assurance without changing the underlying AI model. This is an important source of progress in agentic software engineering: better agents are not only those using more capable AI models, but also the ones operating with better requirements, environments, and evaluators.

These examples define a continuum rather than exceptions to the framework. The requirement gap narrows as authoritative intent becomes more explicit. The model gap shrinks as the relevant world becomes more bounded and stable, or as faithful models of it are learned from data. The evaluation gap shrinks as stronger methods establish conformance within those boundaries. None of these improvements should be mistaken for closure of a different gap.